\documentclass[prb,twocolumn,preprintnumbers,amsmath,superscriptaddress]{revtex4}
\usepackage{bm}
\usepackage{amsmath}
\usepackage[dvips]{graphicx}
\usepackage{color}

\begin{document}
\bibliographystyle{apace}

\title{Negative Hopping Magnetoresistance and Dimensional Crossover\\ 
in Lightly Doped Cuprate Superconductors}

\author{Valeri N. Kotov} 
\affiliation{Department of Physics, Boston University, 590 Commonwealth Avenue, Boston, MA 02215}
\author{Oleg P. Sushkov}
\affiliation{School of Physics, University of New South Wales, Sydney 2052, Australia}
\author{M.~B.~Silva~Neto}
\affiliation{Institut f\"ur Theoretische Physik, Universit\"at
Stuttgart, Pfaffenwaldring 57, 70550, Stuttgart, Germany}
\author{L.~Benfatto}
\affiliation{Centro Studi e Ricerche ``Enrico Fermi'', via Panisperna 89/A,
00184, Rome, Italy}
\affiliation{CNR-INFM and Department of Physics, University of Rome ``La
  Sapienza'', 00185, Rome, Italy}
\author{A. H. Castro Neto}
\affiliation{Department of Physics, Boston University, 590 Commonwealth Avenue, Boston, MA 02215}
\affiliation{Department of Physics, Harvard University, Cambridge, MA 02138}

\begin{abstract}
We show that, due to the weak ferromagnetism of La$_{2-x}$Sr$_x$CuO$_4$, an
external magnetic field leads to a dimensional crossover 2D $\to$  3D 
for the in-plane
transport.  The crossover results in an increase of the hole's localization
length and hence in a dramatic negative magnetoresistance in the variable
range hopping regime.  This mechanism quantitatively explains puzzling
experimental data on the negative magnetoresistance in the N\'eel phase of
La$_{2-x}$Sr$_x$CuO$_4$.
\end{abstract}
\maketitle

\section{Introduction}
The physics of the high-temperature superconducting oxides is determined by
the interplay between the charge and spin degrees of freedom, 
ultimately responsible for the superconductivity itself. A variety
of interesting phenomena exists already at low doping when the oxide layers
are insulating.
In  La$_{2-x}$Sr$_x$CuO$_4$ (LSCO), the
insulating (spin-glass) region corresponds to doping $x<0.055$, with
incommensurate magnetism which exists down to the boundary with the
antiferromagnetic phase (at $x = 0.02$), and even inside the N\'eel region
($x<0.02$). \cite{Matsuda}  A popular point of view favors an explanation
of the incommensurate magnetism based on the tendency of the holes to form
``stripes." \cite{Kivelson}  However, experimental data on variable range
hopping (VRH) (see the review Ref.~\onlinecite{Kastner}), unambiguously indicate
localization of holes for $x<0.055$ and therefore support 
an approach based on a purely magnetic scenario, where a spiral distortion
of the spin background is generated by  localized holes.  The
corresponding theory explains quantitatively the variety of magnetic and
transport data in LSCO. \cite{HCMS,juricic04,SK,JSNMS,LMMS,luscher}

Magnetic phenomena in the low-doping region reflect, in addition to the
Heisenberg exchange, the presence of anisotropies in the spin-spin
interactions, such as Dzyaloshinsky-Moriya (DM) and XY terms.  In the
present paper we consider the N\'eel phase, $x<0.02$.  In this phase the
anisotropies confine the spins to the $(ab)$ plane and fix the direction of
the N\'eel vector to the $\hat{b}$-orthorhombic axis. Moreover, the DM
interaction induces a small out-of-plane spin component that is
ferromagnetic in the plane (weak ferromagnetism) but staggered in the
out-of-plane $\hat{c}$-direction. This component can be easily influenced
by an external magnetic field applied in different directions, as it has
been recently addressed both experimentally
\cite{Thio,Ando,Ono,Gozar,Keimer} and theoretically. \cite{Lara,LS}  For
example, a perpendicular field ($H\parallel\hat{c}$) can cause an alignment
of the out-of-plane moments via a spin-flop transition at a critical field
$H_{f}$, determined by the competition between the DM and inter-layer
Heisenberg exchange (typically $H_{f} \approx 5-7 \ {\mbox T}$). 
\cite{Ando,Gozar,Keimer}  Perhaps most intriguingly, the in-plane
resistivity (along with the $\hat{c}$-axis resistivity) decreases by as
much as $50 \%$ across such a transition. \cite{Thio,Ando}  The magnitude
of the magnetoresistance (MR) shows a rapid increase only below $\approx50$ K 
\cite{Ando} where LSCO exhibits VRH conduction. \cite{Kastner, Lai}
This implies that the MR is accumulated mostly in transitions between localized
states.
 Therefore it is very natural to assume  that the large negative MR is
due to an increase of the hole's localization length as it was 
  suggested in the first experimental paper. \cite{Thio}
From theoretical viewpoint the problem is why the localization length increases
at the spin flop transition.
The first model for the localization length increase,  invoking a three-dimensional 
(3D) VRH  mechanism,  was  proposed in Ref.~\onlinecite{Aharony}.
 However, it is clear now that
except for ultra-low temperatures (that we estimate  to be below $\sim 50
\mbox{mK}$), the VRH conduction at zero magnetic field
 is dominated by two-dimensional (2D) physics.  \cite{Kastner,Lai}
Because of this the 3D picture is not able to describe the most recent and detailed
MR data, as we discuss below.  Experiments are performed typically in the temperature
 range of a few Kelvin and higher where  the out-of-plane
 resistivity anisotropy is  large $\rho_{c}/\rho_{ab} \sim 10^{2}-10^{3}$. \cite{Ando}
 While we ultimately expect that at $T\rightarrow 0$ VRH will become 3D,
 in the temperature range of experimental interest the 2D mechanism
is the relevant one, as is clear from the analysis of the 2D-3D crossover
 temperature and the  fits of the hopping conductivity
 presented in the next section.

 In the present work we demonstrate  that 
 the large MR arises from a change of the effective dimensionality of 
the VRH mechanism with applied field. We support our conclusions
 by detailed comparison with recent experiments on magnetotransport
 which can  be described by our theory with excellent accuracy.    
The main  idea of the present work is that a dimensional crossover
(2D $\to$ 3D) occurs at the spin flop, and this is conceptually and quantitatively different from 
the 3D picture of Ref.~\onlinecite{Aharony}. In particular in our approach
 the increase of the MR (and the localization length)
 is not simply due to the change of the out-of-plane effective
 mass as in Ref.~\onlinecite{Aharony}, but rather arises from a change
 in the shape  of the (localized) wave-functions across the spin-flop.  
In the temperature regime that we keep in mind, 1K and higher, 
  the  change  of the out-of-plane effective
 mass   is   a small, secondary effect (which can  
 manifest itself only at ultra-low temperatures where the full 3D
 VRH mechanism is responsible for transport).
We show that the alignment of the weak ferromagnetic moments in
neighboring planes with the field allows the inter-layer hopping of 
localized holes, which in turn leads to an increase of the hole's in-plane  hopping
probability and thus  negative MR.
The  presence of an inter-layer  hopping channel across
 the spin-flop was  already identified in  Ref.~\onlinecite{Aharony};  
 however our analysis differs in the effects this additional channel
 can produce  in VRH conduction.
 By investigating the evolution of
the hole bound state as a function of  magnetic field and 
temperature, we  find that in various regimes different numbers of
layers are involved in  transport.  In the experimentally relevant temperature
 range the hopping turns out to be quasi-two-dimensional, leading to a negative MR in very good
agreement with the most recent experiments. \cite{Ando,Ono} 

The paper is organized as follows. In Section   II
we analyze the effect of the magnetic field on the dispersion
 of the localized holes, through the inter-layer hopping.
  In Section  III  we present a detailed analysis of
 the change of the hole's wave-function, due to the modified dispersion.
In Sections IV and V we then use the wave-functions to calculate the
 magnetoresistance for  out-of-plane  and in-plane magnetic fields,
 and compare with experiment.  Section VI contains our conclusions.
 
\section{Inter-plane hopping and Spin-flop transition for magnetic field $H
\parallel \hat{c}$}
 First we briefly summarize  previous results
 related to the structure of the hole's 2D bound state at zero field.
 As a starting point, we
consider the hole dynamics in the antiferromagnetic background within the
framework of the $t-t'-t''-J$ model. \cite{SK}  In the absence of the
Coulomb potential $V(r)$ of the Sr ion, a  hole resides, in momentum space,
near the nodal points $(\pm\pi/2,\pm\pi/2)$ and has dispersion
$\epsilon_{\bf k}\approx \frac{\beta_1}{2}k_1^2+\frac{\beta_2}{2}k_2^2$,
where $\beta_1 \approx \beta_2 = \beta \equiv m_{\|}^{-1} \approx 2J$ is
the inverse 2D effective mass appropriate for LSCO \cite{SK} ($m_{\|}
\approx 2m_{e}$ in absolute units).  We measure energies in units of $J=130
\mbox{meV}$, and the lattice spacing is set to unity. 
  Due to $V(r)$, the hole is localized, and its wave function
has the form $\psi(r) \sim e^{-\kappa_0 r}$, corresponding to  binding
energy $\epsilon_0=\beta\kappa_0^{2}/2$. Here the inverse (2D)
localization radius for LSCO is $\kappa_0 \sim 0.3-0.4$, \cite{SK} giving
$\epsilon_0 \approx 10 \mbox{meV}$.  On a perfect square lattice
the bound state is four-fold degenerate: the hole can reside on either up
or down sub-lattices (pseudospin), and it can reside in either of the two pockets
$(\pi/2,\pm\pi/2)$ (flavor). The
orthorhombic distortion lifts the flavor degeneracy due to the presence of
diagonal next-nearest neighbor hopping $t'$. Hence, the  holes occupy only the
pocket $(\pi/2,-\pi/2)$. \cite{LMMS} This is also consistent with the fact
that the spin structure becomes incommensurate along the $\hat{b}$
orthorhombic direction, as seen in neutron scattering for $x< 0.055$.
\cite{Matsuda,LMMS}

Now let us consider  the correction to the 2D dispersion
$\delta\epsilon_{\bf k}$ arising from the inter-layer hopping $t_{\perp}$.
Without account of correlations we have $\delta\epsilon_{\bf
k}=-8t_{\perp}\cos(k_x/2)\cos(k_y/2)\cos(k_z)$, with $\bf{k}$-dependent
$t_{\perp}$, $t_{\perp}=t_c\frac{1}{4}(\cos{k_x}-\cos{k_y})^2$,
where $t_c\sim 50 \mbox{meV}$. \cite{And}  By averaging over the momentum distribution
($k_x,k_y$) in the 2D bound state, we find the effective value of
inter-layer hopping $t_{\perp}\to t_c\kappa_0^2/4 \sim 1 \mbox{meV}$. Since
this is  a crude, order of magnitude estimate,
 below we will use $t_{\perp}$ as a fitting parameter.

The $t-J$ model correlations change $\delta\epsilon_{\bf k}$.  First, the
hopping matrix element should be replaced by $t_{\perp} \to Zt_{\perp}$,
where $Z\approx 0.3$ is the quasiparticle residue.  Second, direct hopping
is allowed only between spins in the same sublattice. In LSCO a spin in a
given plane (e.g. spin ``1" in  Fig.\ 1) interacts with four others in the plane above (and below), 
\cite{Kastner} but at $H=0$ (or $H<H_f$) only out-of-plane hopping in the
$\hat b$ direction is allowed, because this corresponds to ferromagnetic
ordering of spins in neighboring planes (see Fig.\ 1).  However, when a
magnetic field is applied along the $\hat{c}$-axis the spins in the next
layer reverse their signs across the spin-flop transition at $H_f$, so that
for $H > H_{f}$ only hopping in the $\hat{a}$ direction contributes. This
is schematically shown in Fig.\ 1, and reflects in the effective dispersion:
\begin{eqnarray}
\label{zdispersion}
&&\delta \epsilon_{\bf k} = -4Zt_{\bot} \cos(k_{z}) \cos \left(\frac{k_{x} \pm k_{y}}{2}\right),
\\
&& ``-": H < H_{f} \ , \ \ ``+": H > H_{f}, \nonumber 
\end{eqnarray}
where we define the z-direction $z \parallel \hat{c}$. 
Since in the occupied pocket $k_x\approx\pi/2$, $k_y\approx-\pi/2$,
Eq.\ (\ref{zdispersion}) reads
\begin{eqnarray}
\label{dispersion2}
&&\delta \epsilon_{\bf k}^{H<H_{f}} \approx 0, \ \ H<H_{f}\nonumber   \\
&&\delta \epsilon_{\bf k}^{H>H_{f}} =
-4Zt_{\bot} \cos(k_{z}), \ \ H>H_{f}.
\end{eqnarray}
Thus at $H < H_f$ there is no z-dispersion, the coherent dynamics is purely
2D and the VRH in-plane resistivity behaves as $\rho = \rho_{0}
\exp{(T_{0}/T)^{1/3}}$. \cite{Boris} $T_{0}$ is
strongly doping and sample dependent, and the data at zero field,
$x=0.01$ can be fitted by $\rho_{0}= 8\times10^{-6} \Omega{\mbox{cm}}$ and
 $T_0 = 3.6\times10^4 \mbox{K}$ with astonishing accuracy in the range $4{\mbox{K}}<T<50$K,
 as shown in Fig.~\ref{Fig2}(Inset).

 Observe that a
more accurate estimate of the z-dynamics below the spin flop can be
obtained by expanding the in-plane dispersion $\epsilon_{\bf
k}+\delta\epsilon_{\bf k}$ around $(\pi/2,-\pi/2)$ and minimizing the
resulting quadratic form.  This gives a non-zero dispersion in the
z-direction, but the corresponding effective mass is huge,
$M_{\perp}=\beta/[8(Zt_{\perp})^2] \sim 10^4m_e$. Consequently we find that
at temperatures below $T^{*} \sim 2^{-4} \kappa ^{-3}
(m_{\|}/M_{\bot})^{3/2} T_{0}\sim 50 \mbox{mK}$, the VRH is ultimately
three-dimensional, but this regime is irrelevant to present experiments.
\begin{figure}[ht]
\centering
\includegraphics[height=190pt, keepaspectratio=true]{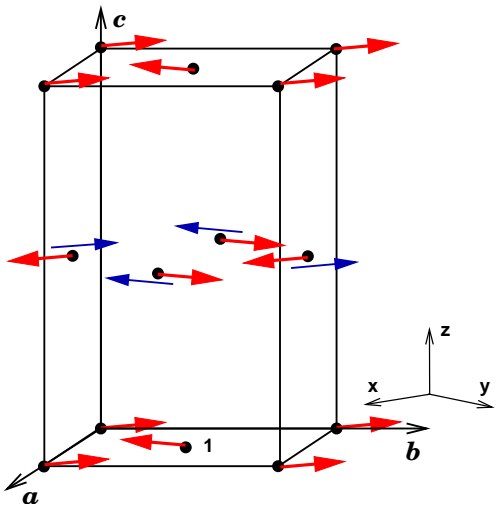}
\caption{(Color online.) Spin structure of  LSCO. Red  arrows correspond to 
  $H<H_{f}$; the blue arrows in the middle layer  show the spin
 reversal for   $H>H_{f}$.}
\label{Fig1}
\end{figure}
\noindent

\section{Evolution of the hole bound state across the spin-flop transition}
To qualitatively understand the effect of the change of dispersion
(\ref{dispersion2}) at $H > H_f$ on the in-plane hole dynamics, we first
estimate the change in the hole binding energy. From (\ref{dispersion2}), after the
spin flop the edge of the continuum ($k_{z}=0$) decreases by $4Zt_{\perp}$,
while the absolute energy of the bound state, to second order in
 the small parameter $Zt_{\perp}/\epsilon_0\ll 1$,
decreases only by amount $\Delta E \sim (Zt_{\perp})^2/\epsilon_0 \ll
t_{\perp}$. The binding energy, which is the magnitude of the difference
between the absolute energy and the continuum limit, changes after the flop
as
\begin{equation}
\label{binding}
\epsilon_0 \to \epsilon \approx \epsilon_{0} - 
4Z t_{\bot} +(Z t_{\bot})^{2}/\epsilon_{0}
\approx \epsilon_{0} -  4Z t_{\bot} \ .
\end{equation}
Within the VRH picture the conduction is proportional to the hole's hopping
probability, which decays exponentially away from the donor site due to the
hole localization. Thus, the decrease of the hole  binding energy 
(\ref{binding}) across the spin flop 
signals an increase of the localization length, and in turn
an increase of the VRH conductivity. 
 This is our central idea that explains the negative MR.

To make this argument more quantitative, we need to compute the change in
the hole's hopping probability across the spin-flop transition.  Let us
enumerate planes by the index $n$ and assume for simplicity that Sr
produces a {\it local} potential $V(r)$ that acts only in the plane
$n=0$. For a shallow level the exact form of the local potential is not
important, \cite{LL} and for simplicity we will  use the $\delta$-function
approximation: $V(r)=-\frac{g}{r}\delta(r-r_0)$, where $r_0$ is assumed to be
 smaller than the localization length, $r_0  \ll 1/\kappa_0$.  The bound
state is described by the wave function $\psi_n(r)$ that depends  on both 
$n$ and $r$. Before the spin flop, $H < H_f$, it obeys the Schr\"{o}dinger
equation:
$\left(-\frac{\beta}{2}\Delta_r-\delta_{n0}\frac{g}{r}\delta(r-r_0)
\right)\psi_n(r)=E_0\psi_n(r)$, where
$E_0=-\epsilon_0=-\beta\kappa_0^2/2$. The solution for $r < r_0$ is given
by $I_0(\kappa_0r)$, and for $r > r_0$ it reads
\begin{eqnarray}
\label{psi0n}
\psi_n&=&0, \ \ \ n\ne 0 \ ,  \nonumber\\
\psi_0&=&\frac{\kappa_0}{\sqrt{\pi}}K_0(\kappa_0r)
 \sim \sqrt{\frac{\kappa_0}{2r}} e^{-\kappa_0r}, \  \kappa_0r \gg 1 \ , \nonumber \\
H& <& H_{f} ,
\end{eqnarray}
where $I_0(r)$ and $K_0(r)$ are the modified Bessel functions of the first and second
kind respectively. For the inverse localization length we obtain
 $\kappa_0=\frac{2e^{-\gamma}}{r_0}\exp{(-\beta/(2g))}$,
 where $\gamma=0.577$ is Euler's constant.
However the exact dependence of  $\kappa_0$ on the parameters of the potential
 is not important, since the energy scale that  $\kappa_0$
 determines, the binding energy $\epsilon_0=\beta\kappa_0^2/2$,
was  extracted from experiment in earlier work, $\epsilon_0 \approx 10 \mbox{meV}$ 
(see Section II).
  
For $H > H_f$ the Schr\"{o}dinger equation becomes
\begin{equation}
\left(-\frac{\beta}{2}\Delta_r +\delta_{n0} V(r) \right)\psi_n
-2Zt_{\perp}\left[\psi_{n+1}+\psi_{n-1}\right] =E\psi_n , 
\end{equation}
with $V(r)=-\frac{g}{r}\delta(r-r_0)$, and $E=-\beta\kappa^2/2$. 
After the Fourier transform $\psi_n(r)=\sum_p\phi_p(r)e^{ipn} \ ,$
we obtain 
\begin{equation}
-\frac{\beta}{2}\Delta_r\phi_p(r) +
V(r)\psi_0(r)=(E+4Zt_{\perp}\cos p)\phi_p(r) \ .
\end{equation}
 By solving this equation we find
\begin{eqnarray}
\kappa/\kappa_0&\approx&
 1+2\left(Zt_{\perp}/\epsilon_0\right)^2 \ ,
\nonumber\\
\phi_p(r)&=&\frac{\kappa_0}{\sqrt{\pi}}
K_0(r\sqrt{\kappa^2-(8Zt_{\perp}/\beta)\cos p} ) \ .
\end{eqnarray}
Thus, as we have already pointed out before Eq.\ (\ref{binding}), after the
spin flop the absolute energy $E$ is shifted only in the second order in
$Zt_{\perp}/\epsilon_0$ and hence this shift can be neglected. Consequently
below we set $\kappa=\kappa_0$.  However, in contrast to Eq.\ (\ref{psi0n}), 
the new wave function,
\begin{eqnarray}
\label{psin}
\psi_n(r)&=&\frac{\kappa_0}{\sqrt{\pi}}\int\frac{dp}{2\pi}e^{ipn}
K_0(\kappa_0r\sqrt{1-(4Zt_{\perp}/\epsilon_0)\cos p} ) , \nonumber \\
H&>&H_{f} ,  
\end{eqnarray}
does not have a simple exponential decay.

\vspace{0.3cm}
\begin{figure}[ht]
\centering
\includegraphics[height=180pt, keepaspectratio=true]{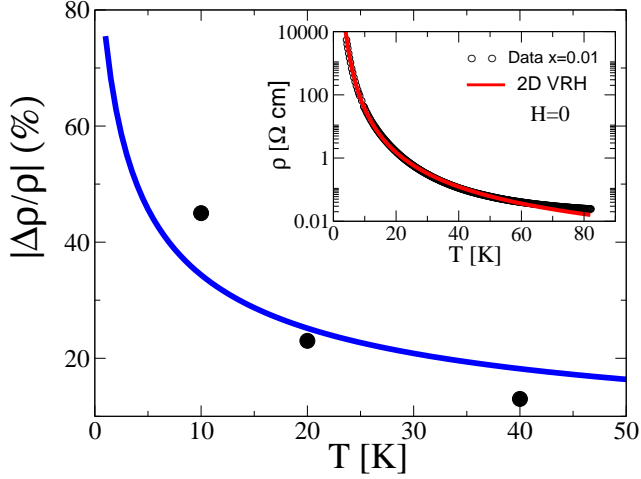}
\caption{(Color online.) The MR jump $|\Delta \rho/\rho|$ (solid, blue line) 
 across the spin-flop transition for $H\|\hat{c}$.
Here $Zt_{\perp}/\epsilon_0 =0.057$, $T_0 = 3.6\times10^4 \mbox{K}$. The circles are
 data from Ref.~\onlinecite{Ando}. Inset: Zero field in-plane resistivity fit to 2D VRH form
(solid, red line) with  $T_0 = 3.6\times10^4 \mbox{K}$.}  
\label{Fig2}
\end{figure}
\noindent

\section{Magnetoresistance across the spin-flop transition for field $H
\parallel \hat{c}$}  To evaluate the MR across the spin flop we compute
now the change of the hole's probability to propagate at the (large) VRH
distance $R_T$: $\kappa_0R_T=\frac{1}{3}\left(\frac{T_0}{T}\right)^{1/3}$.
\cite{Boris}  We can then identify two large-distance regimes for the wave
function (\ref{psin}): (1) Very large distances, $\kappa_0r \gg
\epsilon_0/(2Zt_{\perp})$, and (2) Intermediate large distances,
$\epsilon_0/(2Zt_{\perp}) \gg\kappa_0r \gg 1$. 

 In the first regime the
integral in (\ref{psin}) can be evaluated using the saddle-point
approximation and the wave function is spread over many transverse
channels, $n\sim \sqrt{4\kappa_0 r Zt_{\perp}/\epsilon_0} \gg 1$.  The
probability density at distance $r$, $P(r,t_{\perp})= \sum_n
|\psi_n(r)|^2$, is
\begin{eqnarray}
P(r,t_{\perp})&=&\frac{\kappa_0}{2 r}
\frac{1}{\sqrt{8\pi(Zt_{\perp}/\epsilon_0)\kappa_0 r}}
e^{-2\kappa_0 r\sqrt{1-(4Zt_{\perp}/\epsilon_0)}}, \nonumber \\
 &\mbox{for}&  \  \kappa_0r \gg \epsilon_0/(2Zt_{\perp}) .
\end{eqnarray}
%
In this regime the wave function has a pure exponential decay with a
localization length corresponding to the shift of the binding energy given
in Eq.\ (\ref{binding}). We then find that  
the ratio of conductivities after ($H > H_f$) and before ($H < H_f$) the
flop is
\begin{equation}
\label{ult}
\frac{\sigma_{H > H_f}}{\sigma_{H < H_f}}=
\frac{P(R_T,t_{\perp})}{P(R_T,t_{\perp}=0)}\sim
\exp\left\{\frac{4Zt_{\perp}}{3\epsilon_0}
\left(\frac{T_0}{T}\right)^{1/3}\right\} \ .
\end{equation}
Here $P(R_T,t_{\perp}=0)=|\psi_0|^{2}$, with $\psi_0$ from
Eq.(\ref{psi0n}). This result is valid when $\kappa_0 R_T\sim
(T_0/T)^{1/3}\gg \epsilon_0/(2Zt_\perp)$, which happens in practice when the
temperature is  below $1\mbox{K}$.  In this case $\sigma_{H >
H_f}/\sigma_{H < H_f}\gg 1$, i.e. the corresponding MR is large: $|\Delta
\rho/\rho \equiv (\rho_{H > H_f}/\rho_{H < H_f}) -1| \approx 1$, and a full
2D $\to$ 3D crossover is expected in the spin flop.  Notice that the MR is
always negative, $\Delta \rho/\rho <0$.

In the Intermediate large distances regime, $\epsilon_0/(2Zt_{\perp})
\gg\kappa_0r \gg 1$, the integral in (\ref{psin}) can be evaluated by direct
expansion in powers of $\kappa_0rZt_{\perp}/\epsilon_0$.  Only the layers
$n=0,\pm 1$ contribute in this case, $P(r,t_\perp)= \sum_{n=0,\pm
1}|\psi_n|^2$, leading to
\begin{eqnarray}
\label{prs}
P(r,t_\perp)&=&  
\frac{\kappa_0}{2 r}e^{-2\kappa_0 r}
\left(1+4(\kappa_0 r)^2\frac{(Zt_{\perp})^2}{\epsilon^2_0}\right), \\
 & \mbox{for} &  \ \epsilon_0/(2Zt_{\perp}) \gg\kappa_0r \gg 1 . \nonumber
\end{eqnarray}
From (\ref{prs}) we obtain ($r\rightarrow R_T$)   across the spin flop
\begin{eqnarray}
\label{slxs}
\frac{\sigma_{H > H_f}}{\sigma_{H < H_f}}
=1+\frac{4}{9}\left(\frac{T_0}{T}\right)^{2/3}\frac{(Zt_{\perp})^2}{\epsilon^2_0} \ .
\end{eqnarray}
This formula corresponds to a crossover from a pure 2D case to an
``intermediate dimension" (three transverse channels), and it is justified
when $(\sigma_{H > H_f}/\sigma_{H < H_f}) -1 \ll 1$.
We have also performed a full numerical evaluation of $P(r,t_\perp)$ and of
the MR with $\psi_n(r)$ from Eq.\ (\ref{psin}), since the above
considerations are based on asymptotic  behavior.
The exact numerical form was used in both Figures \ref{Fig2} and \ref{Fig3}
below. We have found
quite clearly that indeed in the temperature range where experimental data
are available, $T > 10 \mbox{K}$, the intermediate asymptotic formula (\ref{slxs})
is the relevant one, since the MR is still relatively  small there,
e.g. $|\Delta \rho/\rho| \approx 0.35, T = 10 \mbox{K}$.
 However as the temperature is lowered, $T < 10 \mbox{K}$,
 the MR increases and  the system enters a crossover region between the asymptotic
 formulas Eq.~(\ref{slxs})  and  Eq.~(\ref{ult}), 
 which requires  the use of the exact wave-functions.
  The calculated MR is plotted in Fig.~\ref{Fig2} and we observe 
that the fitted value $Zt_{\perp}/\epsilon_0=0.057$ agrees quite well with
the  estimate $Zt_{\perp}/\epsilon_0 \sim 0.03$ from band structure
 calculations. \cite{And}

\vspace{0.4cm}
\begin{figure}[ht]
\centering
\includegraphics[height=185pt,keepaspectratio=true]{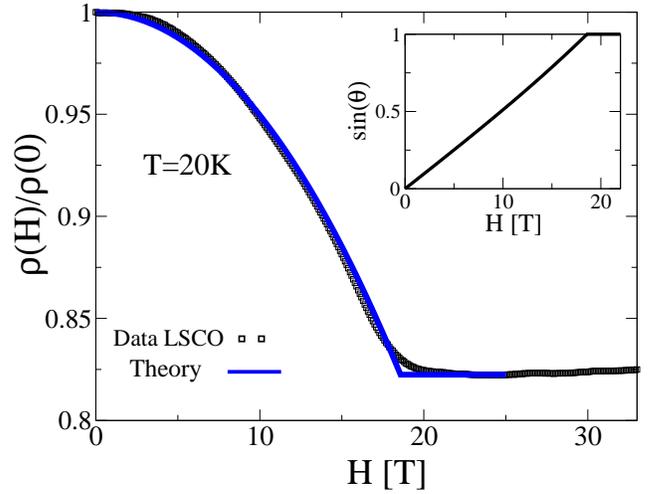}
\caption{(Color online.) The MR $\rho(H)/\rho(0)$ (solid, blue line)
 for in-plane field $H\|\hat{b}$, with $Zt_{\perp}/\epsilon_0 = 0.047$.
The squares are data from Ref.~\onlinecite{Ono}. Inset: Evolution of the angle
 (\ref{Angle}) with field.} 
\label{Fig3}
\end{figure}
\noindent

\section{Magnetoresistance for in-plane field $H \parallel \hat{b}$} In
this case, due to the alignment of the DM-induced moments with the magnetic
field, the spins rotate in the $\hat{b}-\hat{c}$ plane (in opposite
directions on the two sub-lattices), \cite{Thio,Ono,Gozar,Lara,LS,ML} and
align completely along $\hat{c}$ at a field $H_{c2}$. Thus, once  we
introduce the angle $\theta(H)$ that the spins make with the
$\hat{b}$-direction, our previous calculations leading to Eq.\ (\ref{slxs})
remain valid with the replacement $t_{\bot} \rightarrow t_{\bot}
\sin{\theta}$. Observe that while for the undoped LCO an intermediate
in-plane spin flop is expected at $H_{c1}<H_{c2}$, \cite{Lara,LS} in LSCO
the doped holes contribute to enhance the $\hat{b}$-axis spin
susceptibility and then to confine the spins in the $\hat{b}-\hat{c}$
plane, as it has been discussed recently in Ref.~\onlinecite{ML}.  We can then
write the field dependence of $\sin\theta$ between $H=0$ and $H=H_{c2}$ as
\begin{equation}
\sin\theta(H)= \frac{H D/\sigma_0}
{\Delta_{out}^2+4\eta-(1- x \;  \chi_{imp})H^2}.
\label{Angle}
\end{equation}
In the above expression $D$ is the DM anisotropy, $\eta=2JJ_\perp$
is the inter-layer exchange, $\Delta_{out}$ is the out-of-plane (or
XY) anisotropy gap, $\sigma_0$ is the staggered order parameter, $x$
is the doping and $\chi_{imp}$ is a dimensionless measure of the
holes-induced spin susceptibility. Here $H$ is measured in units of $
g_s^b\mu_B H$, with $g_s^b=2.1$ and $\mu_B$ is the Bohr magneton. The
parameter values can be extracted from the experiments, 
\cite{Lara,ML} and for $x=0.01$ we take $D=2.16$ meV, $\eta=1$ (meV)$^2$,
$\Delta_{out}=3.2$ meV, $\chi_{imp}=80$, and $\sigma_0=0.36$, which gives
$H_{c2}\approx 19$ T. \cite{Ono} Using $Zt_\perp/\epsilon_0=0.047$ as the
only fitting parameter, we find a remarkable agreement with the experimental
data of Ref.~\onlinecite{Ono} at $T=20$ K, as shown in Fig.~\ref{Fig3}.

 We also note that in Oxygen-doped compounds the contribution of the localized holes to the
longitudinal susceptibility is much smaller, with $\chi_{imp}\sim 1$.
\cite{ML} As a consequence, one expects to observe at a field $H_{c1}\sim
10$ T an intermediate flop which reflects in a kink in both the
$\sin\theta(H)$ curve and in the MR curve, as it has been measured indeed
in the earlier transport measurements in La$_{2}$CuO$_{4+y}$. \cite{Thio}

\section{Discussion and conclusions}
In conclusion, we have shown that in the strongly localized VRH regime for
$T<50 \mbox{K}$, the in-plane MR is sensitive to the inter-layer hopping $t_{\bot}$
because an external magnetic field effectively changes the dimensionality
of the problem, making the hopping quasi-2D. The MR reflects the physics of the spin-flop and
is always negative; its value as well as  temperature and field  dependence are in excellent 
quantitative agreement with recent experiments in LSCO at $x=0.01$. \cite{Ando,Ono} 
 Orbital effects, typically causing positive MR, \cite{Boris} are negligible at this
small  doping because the hole's localization length $1/\kappa \sim 2-3$ is much smaller
than the magnetic length at any reasonable fields.

 Finally we comment that unlike LSCO where the role of magnetic anisotropies is very 
well established, in insulating  YBa$_{2}$Cu$_3$O$_{6+x}$ (YBCO) such anisotropies
 are expected to be very weak, which is related to the absence of strong localization 
in this material. \cite{Sun} 
Spin-related effects in the in-plane MR are absent for field in the $\hat{c}$ direction,
while the in-plane field MR remains  small and appears  to be due to 
the dynamics of holes that are very weakly influenced by
 disorder. \cite{AndoYBCO} The complete understanding of
 these phenomena in YBCO remains an open issue although it is clear that magnetotransport is
 not dominated by the local spin physics.

\acknowledgments
We are grateful to D. K. Campbell, O. K. Andersen, O. Jepsen, A. Lavrov,
and Y. Ando for stimulating discussions. O.P.S. gratefully  acknowledges
support from the Alexander von Humboldt Foundation, and the hospitality of
the Max-Planck-Institute 
for Solid State Research Stuttgart 
and Leipzig University. 
V.N.K. is supported by Boston University.  A.H.C.N. is
supported through NSF grant DMR-0343790.


\begin{thebibliography}{99}

\bibitem{Matsuda} M. Matsuda,     M. Fujita,  K. Yamada,
    R. J. Birgeneau,  M. A. Kastner,     H. Hiraka,  Y. Endoh,
    S. Wakimoto,  and G. Shirane,  
 Phys. Rev. B {\bf 62}, 9148 (2000);
 M. Matsuda,  M. Fujita,  K. Yamada,
R. J. Birgeneau,  Y. Endoh, and  G. Shirane, Phys. Rev. B  {\bf 65}, 134515 (2002). 

\bibitem{Kivelson} S. A. Kivelson,  I. P. Bindloss,
E. Fradkin, V. Oganesyan, J. M. Tranquada, A. Kapitulnik,
and C. Howald, Rev. Mod. Phys. {\bf 75},
 1201 (2003).

\bibitem{Kastner} M. A. Kastner,  R. J. Birgeneau, 
G. Shirane, and Y. Endoh, Rev. Mod. Phys. {\bf 70}, 897
(1998).

\bibitem{HCMS}
  N. Hasselmann, A. H. Castro Neto, and C. Morais Smith,
Phys. Rev. B {\bf 69}, 014424 (2004).

\bibitem{juricic04}  V. Juricic,  L. Benfatto, A. O. Caldeira, and C. Morais Smith,  
Phys. Rev. Lett. {\bf 92}, 137202 (2004).

\bibitem{SK}
 O. P. Sushkov and V. N. Kotov, Phys. Rev. Lett. {\bf 94}, 097005 (2005);
 V. N. Kotov and O. P. Sushkov, Phys. Rev. B {\bf 72}, 184519 (2005).

\bibitem{JSNMS}  V. Juricic, M. B. Silva Neto, and C. Morais Smith,
Phys. Rev. Lett. {\bf 96}, 077004 (2006).

\bibitem{LMMS} A. Luscher,  Gr. Misguich, A. I. Milstein, and O. P. Sushkov, 
 Phys. Rev. B {\bf 73}, 085122 (2006).

\bibitem{luscher} A. Luscher, A. I. Milstein, and O. P. Sushkov, 
Phys. Rev. Lett. {\bf 98}, 037001 (2007).

\bibitem{Thio} T. Thio,  T. R. Thurston, N. W. Preyer,
P. J. Picone, M. A. Kastner,  H. P. Jenssen,  D. R. Gabbe,
    C. Y. Chen,  R. J. Birgeneau, and  A. Aharony,  
  Phys. Rev. B {\bf 38}, 905 (1988);  T. Thio,
 C. Y. Chen, B. S. Freer, D. R. Gabbe, H. P. Jenssen, M. A. Kastner, P. J. Picone,
 N. W. Preyer, and R. J. Birgeneau,
  Phys. Rev. B  {\bf 41}, 231 (1990).

\bibitem{Ando} Y. Ando, A. N. Lavrov, and S. Komiya, Phys. Rev. Lett. {\bf
90}, 247003 (2003).

\bibitem{Ono} S. Ono,  S. Komiya, A. N. Lavrov,  Y. Ando,
F. F. Balakirev, J. B. Betts, and G. S. Boebinger,  
  Phys. Rev. B {\bf 70}, 184527 (2004).

\bibitem{Gozar} A. Gozar,  B. S. Dennis, G. Blumberg, S. Komiya,  and Y. Ando,
 Phys. Rev. Lett. {\bf 93}, 027001 (2004).

\bibitem{Keimer} M. Reehuis,  
C. Ulrich, K. Prokes, A. Gozar, G. Blumberg, S. Komiya, Y. Ando,
 P. Pattison, and B. Keimer,  Phys. Rev. B {\bf 73}, 144513 (2006).

\bibitem{Lara} L. Benfatto and M. B. Silva Neto, Phys. Rev. B {\bf 74},
 024415 (2006);
 L. Benfatto,  M. B. Silva Neto, A. Gozar,
 B. S. Dennis, G. Blumberg, L. L. Miller, S. Komiya, and Y. Ando, 
Phys. Rev. B {\bf 74}, 024416 (2006).

\bibitem{LS} A. Luscher and O. P. Sushkov, Phys. Rev. B {\bf 74}, 064412
 (2006), and cited references.

\bibitem{Lai} E. Lai and R. J. Gooding,  Phys. Rev. B {\bf  57}, 1498 (1998).

\bibitem{Aharony} L. Shekhtman, I. Ya. Korenblit, and A. Aharony,
Phys. Rev. B {\bf 49}, 7080 (1994).

\bibitem{And} O. K. Andersen, private communication.

\bibitem{Boris} B. I. Shklovskii and A. L. Efros, {\it Electronic
 Properties of Doped Semiconductors}, (Springer-Verlag, Berlin, 1984).

\bibitem{LL} L.~D.~Landau and E.~M.~Lifshitz, {\it Quantum Mechanics},
(Pergamon Press,  Oxford, New York, 1965).

\bibitem{ML}
M. B. Silva~Neto and L. Benfatto,  Phys. Rev. B {\bf 75}, 140501(R) (2007).

\bibitem{Sun} X. F. Sun, K. Segawa, and Y. Ando, Phys. Rev. B {\bf 72}, 100502(R) (2005).

\bibitem{AndoYBCO} Y. Ando, A. N. Lavrov, and K. Segawa, 
Phys. Rev. Lett. {\bf 83}, 2813 (1999);  
A. N. Lavrov,  Y. Ando, K. Segawa, and J. Takeya, Phys. Rev. Lett. {\bf 83}, 1419 (1999).

\end{thebibliography}
\end{document}